\documentclass[final,1p,times]{elsarticle}
\journal{arXiv}

\usepackage{graphics,amssymb,multirow,booktabs,amsmath,hyperref,epsfig,color,eurosym, amstext}

\usepackage{graphicx}
\usepackage{lscape}
\graphicspath{{figure/}}

\begin{document}
\begin{frontmatter}

\title{Meet me in the middle: \\ The reunification of Germany's research network}

\author[inst1]{Bogang Jun}
\author[inst1]{Fl\'{a}vio L. Pinheiro}
\author[inst2,inst3] {Tobias Buchmann}
\author[inst4]{Seung-kyu Yi}
\author[inst1]{C{\'e}sar A. Hidalgo\corref{cor1}}
\cortext[cor1]{\emph{Email address}: hidalgo@mit.edu}

\address[inst1]{Collective Learning Group, MIT Media Lab, Massachusetts Institute of Technology, Cambridge, MA 02139, USA}
\address[inst2]{Chair for Innovation Economics, University of Hohenheim, D-70593 Stuttgart, Germany}
\address[inst3]{Center for Solar Energy and Hydrogen Research Baden-W\"urttemberg (ZSW), D-70565 Stuttgart, Germany}
\address[inst4]{Technology Management, Economics and Policy Program, Seoul National University, Seoul, South Korea}

\begin{abstract}
In 1990, Germany began the reunification of two separate research systems. In this study, we explore the factors predicting the East-West integration of academic fields by examining the evolution of Germany's co-authorship network between 1974 and 2014. We find that the unification of the German research network accelerated rapidly during the 1990s, but then stagnated at an intermediate level of integration. We then study the integration of the 20 largest academic fields (by number of publications prior to 1990), finding an inverted U-shaped relationship between each field's East or West "dominance" (a measure of the East-West concentration of a field's scholarly output prior to 1990) and the fields' subsequent level of integration. We checked for the robustness of these results by running Monte Carlo simulations and a differences-in-differences analysis. Both methods confirmed that fields that were dominated by either West or East Germany prior to reunification integrated less than those whose output was balanced. Finally, we explored the origins of this inverted U-shaped relationship by considering the tendency of scholars from a given field to collaborate with scholars from similarly productive regions. These results shed light on the mechanisms governing the reintegration of research networks that were separated by institutions.
\end{abstract}


\end{frontmatter}

\section{Introduction}
History contains many examples of two independent states forming a new unified entity. Take, for instance, the unification of Spain in the fifteenth century and Italy in the nineteenth century, the creation of Yugoslavia in 1918, the unification of Vietnam in 1975, the formation of the United Arab Emirates in 1971 or, more recently, the reunification of Germany in 1990. Among these examples, the reunification of Germany, which is associated with the final years of the Cold War, is somewhat unique, since it provides an opportunity to understand the consequences of the unification of two states with separate research systems, and institutions, but a common language and geography.

According to \cite{Wrobel2010}, the reunification of Germany through the so-called Two-Plus-Four Treaty of 1990 was politically and economically beneficial to the country in the sense that it recovered its sovereignty and became the strongest economy in the EU. Nevertheless, the integration of the two blocks was neither easy nor cheap.

Since 1990, the German government has spent \euro270 billion on a reunification process that is aimed primarily at equalizing standards of living between the East and West; however, despite such immense expenditure, in 2014 the nominal per capita GDP of East Germany was two-thirds that of West Germany (\euro24,324 versus \euro36,280) \citep{BMWi2015}. Furthermore, East Germany has experienced a significantly higher unemployment rate than West Germany.

But what factors can explain the prevalence of the East-West German gap? One possibility, according to \cite{Bach2000}, is that the reunification process is governed not just by formal institutions, but also by organic social and economic forces, like those governing the formation of social networks. Social network formation is ruled by mechanisms, such as triadic closure, homophily, and shared social foci \citep{Hidalgo2016, granovetter1973strength, Borgatti2009, rainie2012networked, christakis2009connected}, which cannot be easily modified through government intervention. According to these theories, we should expect German reunification to be a relatively slow process constrained by pre-existing social structures and the mechanisms known to rule the formation of social and professional networks.

In  principle, the unification of two previously independent states affects multiple social and economic outcomes. Here, we focus on just one of them: the reunification of Germany's research-collaboration network. To explore this reunification we conducted a longitudinal analysis of this network using pairs of NUTS 3 regions (small geographical areas that divide Germany into 429 regions) as nodes, and measured the volume of collaboration (co-authorships) between scholars from East and West Germany for each pair of NUTS 3 regions as links. 

We chose to focus on co-authorship networks because these are considered to be good proxies for social and professional relationships \citep{Newman2001, Newman12001, Newman22001}. According to \cite{Storper2004}, deep social interactions facilitate learning, trust, and the ability to communicate complex ideas, all of which are prerequisites for effective scholarly collaborations.

In this paper, we specifically explore the conditions that led to the successful reunification of academic fields in Germany by examining the evolution of the co-authorship network connecting scholars from East and West Germany between 1974 and 2014. To begin, we measure the unification of this network by studying the evolution of its modularity (a method to estimate the compartmentalization of a network). Our results show that the reunification of Germany's research network initially occurred rapidly during the 1990s, with many new collaborations between East- and West-German institutions, but has stagnated since then, remaining at an intermediate level of integration.  

Next, we examined the 20 largest academic fields (which were determined by analyzing the number of publications produced in each field prior to reunification) and studied whether fields dominated by East or West Germany prior to reunification featured more successful integration of their co-authorship networks. Consequently, we found that neither West- nor East-German-dominated fields integrated most successfully. The fields that integrated the most  boasted a more balanced output between the two sides prior to reunification. This finding is supported by our differences-in-differences (DID) analysis and Monte Carlo simulations. The latter were used to test whether  the balance of field, and not just its size or spatial distribution, was a significant predictor of subsequent integration. Finally, we explored why the balance of a field should be associated with its subsequent integration by checking the tendency of scholars to co-author papers with other scholars from similarly ranked regions. Using numerical simulations, we found that, when this tendency was absent, the U-shaped relationship disappeared, suggesting that the homophily observed in these mixing patterns contributes to the different speeds of reunification between fields with balanced and unbalanced outputs.

\section{Historical background of the reunification of the German research system}

Germany's research system was divided after the Second World War; however, in 1990 a reunification process suddenly began. After Germany lost the Second World War, the country was occupied by allied forces and divided into four military occupation zones, which were divided between the United States, the Soviet Union, France, and the United Kingdom \citep{ardagh1988germany}. Later, as the clash between communism and capitalism manifested into bitter competition between the United States and the Soviet Union, both countries realized that they needed "their" Germany to represent an example of European economic recovery and, by extension, their form of governance \citep{gareau1961morgenthau}.

In 1949, the Soviet Union took steps to consolidate its influence in Europe and transformed its zone of occupation into the German Democratic Republic (GDR). By this time the GDR already possessed a centrally planned economy and a centralized research system. In the same year, the three occupation zones occupied by the United States, the United Kingdom, and France were unified into the Federal Republic of Germany (FRG), and a social market economy and a federal system that included research and education were introduced. During the 1950s, an increasing number of people seeking a better future migrated from East to West Germany; however, this border-crossing movement was eventually halted by the socialist regime with the construction of tough border controls and the construction of the Berlin Wall in 1961. While the socialist economic and scientific system achieved some success, East Germany never reached the strength of West Germany, and after almost 30 years of physical separation, it collapsed in 1989 \citep{pence2008socialist}. During this momentous year, the GDR celebrated its 40th anniversary on October 7th, and on November 9th, only one month later, a peaceful revolution occurred, which resulted in East Germany unexpectedly opening its borders and allowing its citizens to enter West Berlin and West Germany. These developments led to German reunification in 1990 \citep{weber2002kleine}. The above shows that the reunification process occurred somewhat suddenly, and was not greatly anticipated by economic or scientific actors.

In the aftermath, a treaty concerning the economic, monetary, and social union was adopted. This facilitated economic reunification and served as a master plan for the introduction of the social market economy in the former GDR. However, the process of convergence did not occur in the expected manner. East Germany had inefficient industry, obsolete equipment, and most of its products were not competitive in the international market; thus, former East German regions experienced the liquidation of many businesses, high rates of unemployment, and the transfer of skilled labor to Western Germany. Although in 1991 the government implemented a support program for East German regions called "Gemeinschaftswerk Aufschwung Ost," which was designed to foster the economic recovery of the East. 

Article 38 of the 1990 German Unification Treaty stipulates the creation of an integrated German R\&D system that will perform as well as the old West German system. This shows that elements of the GDR system that performed well should have been preserved. In July 1990, West and East German research ministers agreed on the basic elements of a unified research system: "It is our aim to create an integrated research system in a unified Germany. Moreover, the incorporation of the institutions combined under the East German Academy of Sciences into this research scene will be a central task" \citep{BMFT}. As evidenced by the rapid reduction in R\&D personnel, these goals were not realized in the short term, and part of the research structure has since been dissolved.

Between 1989 and 1993, the number of R\&D staff in East Germany decreased to a level as low as 30\% of its initial figure. Indeed, the level of employment in R\&D relative to the total number of people employed in East Germany quickly fell below 50\% of its West German equivalent whereas, in 1989, there were no major differences in the ratios for the two states. Most existing universities were preserved; however, research at universities was reduced by 30\%, meaning East Germany accounted for, at best, 10\% of the total research capacity in the German higher education system, despite the region accommodating 20\% of the population. 

In the GDR, non-university R\&D institutions dominated; in particular, the East German Academy of Sciences, the Academy of Agricultural Sciences, and the Building Academy. For example, the Academy of Sciences' mission was to conduct basic research and use the results in applied research. This organization's important role was substantiated by the fact that R\&D-intensive industries, such as the chemical industry and electronics and mechanical engineering, had less R\&D personnel than equivalent West German sectors; consequently, academy scientists often performed contract research for different industries. However, once the total number of R\&D personnel is taken into account, the performance of the East German R\&D system is found to be significantly lower when compared to that of West Germany \citep{Meske1993}.

Numerous programs were implemented in an attempt to accelerate the slow convergence between the two systems. According to \cite{Gunther2010} and the Federal Ministry of Education and Research \citep{BMBF} of Germany, the period after reunification in 1990 can be divided into three policy regimes: (i) the reconstruction of the old system between 1990 and 1997, (ii) the introduction of a new system between 1998 and 2006, and (iii) its respective stabilization since 2007. During the initial years, these programs mainly aimed at restructuring the scientific landscape of Eastern Germany. Only in the second and, in particular, the third phase was greater emphasis placed on collaboration and knowledge transfer between east and west. These observations support the design of our study as a kind of natural experiment of the integration of two formerly separated networks. 

Indeed, until the end of the 1990s, innovation politics focused on fostering firms' R\&D activities and innovative entrepreneurship in East Germany. Since then, however, a re-adjustment of the policy approach based on knowledge transfer and network formation has been implemented \citep{eickelpasch2015forschung}. One good example of this new policy approach is the InnoRegio program, which operated from 1999 to 2006, as this aimed to boost competition among 23 networks of firms and research facilities. This prominent policy was based on the idea that innovation is not driven by a single individual or single Schumpeterian entrepreneur, but rather by networks consisting of various participants, organizations, and institutions. Its main objective was to improve the transfer of knowledge and technology between East German regions by building networks that had a special focus on SMEs (since the main actors in East Germany were SMEs), rather than on the large companies that had been the main actors in West Germany. In another development, the Innovationsforen program was introduced, which also supported the early phases of innovation networks in East Germany in an attempt to strengthen the development of a thematic focus and collaborative relations. Another example is a program called Entrepreneurial Regions (Unternehmen Region), which was implemented in 2004. This policy also aimed to form strong interlinked regions to promote a free exchange of knowledge. Key elements of this renewed strategy are lateral thinking, cooperation, strategic planning, and entrepreneurial action. To implement this new strategy, several initiatives have been deployed to promote national, inter-, trans-, and multidisciplinary cooperation between partners and encourage openness and transparency. Under these initiatives, networks are to be formed across East Germany that feature one or more partners from West Germany but which include a project leader from the east. Our expectation is that these initiatives have already triggered many new collaborative ties between East and West German researchers.

\section{Data}

As material for this study, we used publication data from the journals listed in the Science Citation Index (SCI) of Web of Science (WoS). We considered all article types (journal articles, conference proceedings, reviews, letters, news, and book reviews) published between 1972 and 2014. Our final data contain all papers for which at least one author had a German address. Further, we also collected article IDs, author's addresses, the fields of study, and the institutions to which the authors belong. In total, our dataset consists of 2,897,527 papers. Since we were focusing on papers that connect regions, we generally used papers that included at least two authors based in two different regions (constituting 1,371,639 of the total number), but when we calculated the level of dominance of a region in certain field (we will explain the measure of dominance later in this paper), we used all data, including papers with a single author. Moreover, we noticed a large, unexpected jump in the number of publications between 1973 and 1974 and, consequently, we decided to omit data from 1972 and 1973. 

\begin{figure}[ht]
  \centering
  \includegraphics[width=0.8\textwidth]{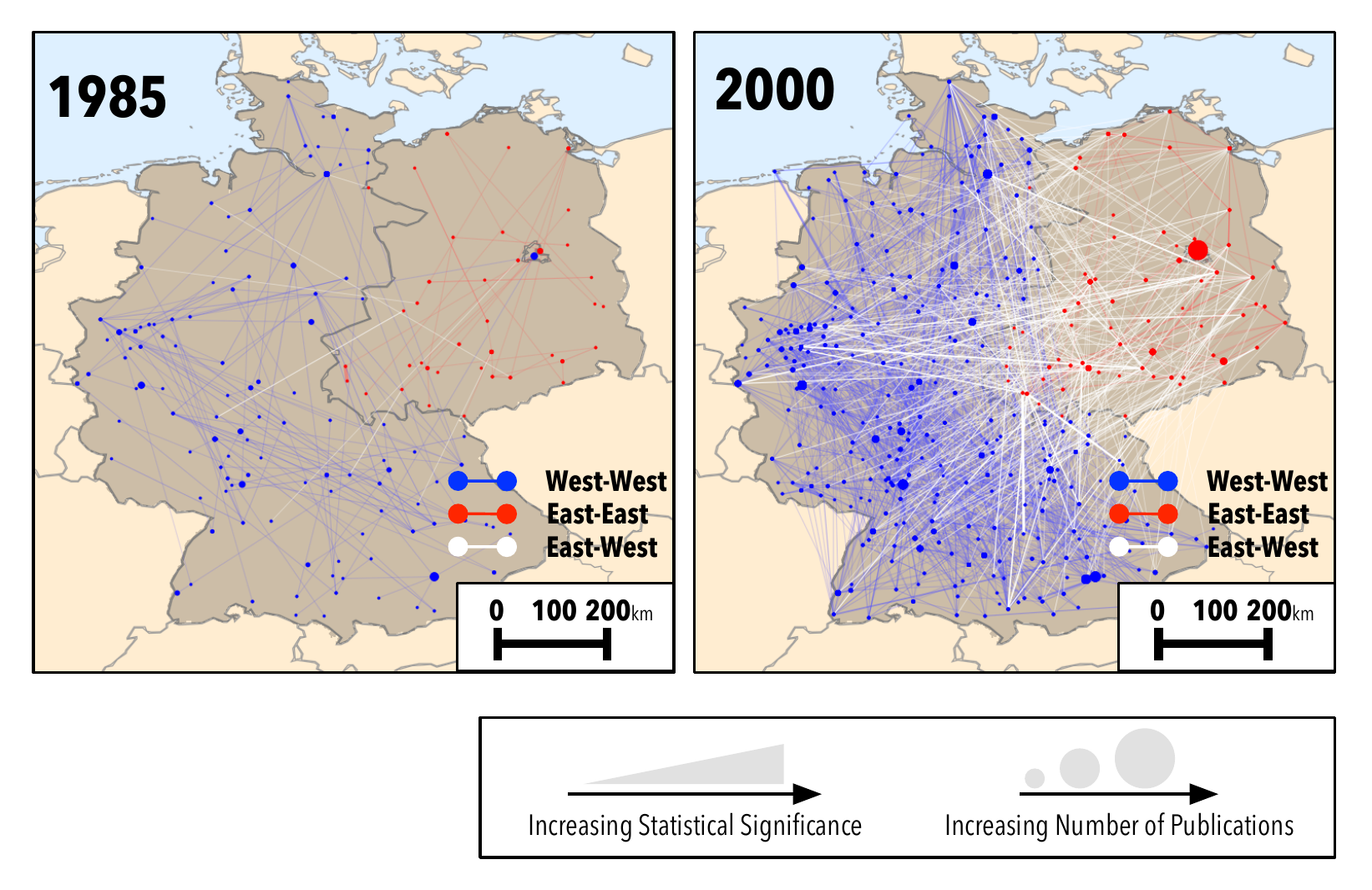}
  \caption{Graphical representation of the collaboration network overlaying a map of Germany. White lines represent collaborations between West and East, while blue and red lines represent collaborations within West and East, respectively. The thickness of the links is proportional to the level of statistical significance, whereas node sizes are proportional to the number of publications of each region.}
  \label{dataSummary1}
\end{figure}

\begin{figure}[!b]
  \centering
  \includegraphics[width=0.8\textwidth]{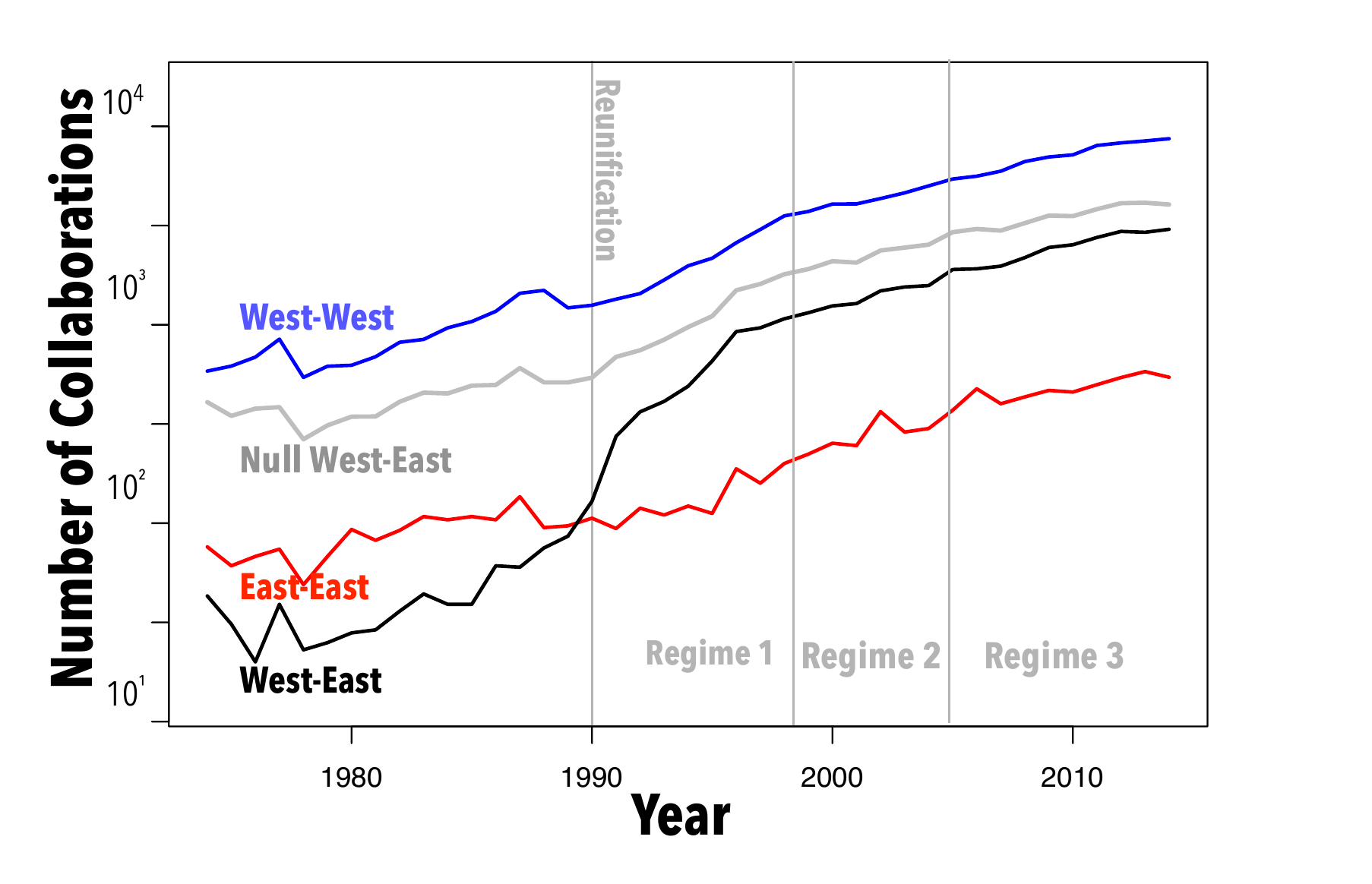}
  \caption{Blue and red lines represent the number of collaborations within West and East, respectively, while the black line represents the number of collaborations between West and East. The gray line relates to the number of collaborations between East and West from the null model.}
  \label{dataSummary2}
\end{figure}

In our analysis, two regions were connected when authors from those regions had published a research article together. Therefore, in our collaboration network, nodes represent regions and links correspond to the number of papers that include authors from each region, as depicted in Figure~\ref{dataSummary1}. Throughout the remainder of this paper, the regions that belonged to former GDR regions are labeled as "East Germany," while former FRG regions are labeled as "West Germany." Considering that our research interest concerns collaborations between East and West, we ignored other collaborations with researchers from foreign regions. For example, if a paper was co-authored by three authors, with addresses in West Germany, East Germany, and the US; we recorded the collaboration between West and East Germany and ignored the collaboration between West Germany and the US and between East Germany and the US. We agree that the role of such third parties is interesting, but it goes beyond the scope of our current study. During the time period analyzed, we found that the number of German regions linked with NUTS 3 regions in at least one paper increased from 169 (39\%) in 1974 to 391 (91\%) in 2014. Additionally, the number of pairs of German regions with at least one co-authored paper also increased during this period, from 703 (0.7\% of all possible pairs) in 1974 to 12,240 (13.3\% of all possible pairs) in 2014. 

Figure~\ref{dataSummary1} shows the network of relationships between German regions that existed in 1985 and in 2000. White lines represent the links between East and West, while red and blue lines represent the collaboration links within East and West, respectively. As can be seen from the figure, there are few white lines (East-West collaborations) in 1985, but many in 2000. 

To depict the number of collaborations over time, we plot Figure~\ref{dataSummary2}, showing the evolution of the collaborations within East Germany (red) and West Germany (blue), and between West and East Germany (black) from 1974 to 2014. Additionally, the gray line shows the number of collaborations between West and East Germany estimated by a null model created through Monte Carlo simulations. In this null model, real data is used to connect each region to the other regions (i.e., each node's degree remains consistent between the real network and the null model), but collaborations are assigned at random, subject to this constraint (further information concerning the null model is provided in Appendix B). By comparing the gray and black lines we can determine if the volume of collaboration is higher or lower than our estimations. The true figure obtained shows that the volume of East-West collaborations was drastically lower than that of the null model prior to 1990 and then increased between 1990 and 1997 (approaching the value of the null model); since then, it has remained below the null model, implying that the maximum potential level of collaboration between East and West Germany was not achieved.

\section{Methods and results}

\subsection{Measuring the speed of network unification using modularity}
Since our aim is to analyze the relationship between the dominance of a field and the speed of unification, we measure the speed of unification and the level of dominance for each field. First, to measure the unification of the German research network, we employ a network-analysis tool called modularity \citep{Borgatti2009, Hidalgo2016, Newman2003, Fortunato2010}. Modularity \citep{Fortunato2010, Fortunato2011, Newman2006} is a tool for the detection and characterization of communities in networks (groups of densely interconnected nodes with relatively few connections to other groups). Community detection is a popular topic among network scientists since communities may correspond to social or functional units in networks \citep{Newman2006}. 

We estimate modularity following the seminal work of \cite{Newman2006}. \cite{Newman2006}'s theory is that "if the number of links between groups is significantly less than we expect by chance, or equivalent if the number of within groups is significantly more," we can say that a network exhibits a community structure. Consequently, modularity Q, is "the number of edges falling within groups minus the expected number in an equivalent network with edges placed at random." Formally, modularity Q is defined as: : 

\begin{equation}
\mathcal{Q} = \frac{1}{4} \sum_{ij} \left( A_{ij} - \frac{k_i k_j}{2 m} \right) \left( s_i s_j + 1 				\right)
\label{Moduality}
\end{equation}

where $A_{ij}$ is the adjacency matrix of the network; $k_{i}$ and $k_{i}$ are the degree of nodes $i$ and $j$, respectively; and $m$ is the total number of links, which is equal to $\frac{1}{2}\sum_{i}k_{i}$. Supposing two groups exist in a network, $s_{i}$ is +1 if node $i$ belongs to group 1, and $s_{j}$ is -1 if node $j$ belongs to group 2. 

\begin{figure}[!t]
  \centering
  \includegraphics[width=0.8\textwidth]{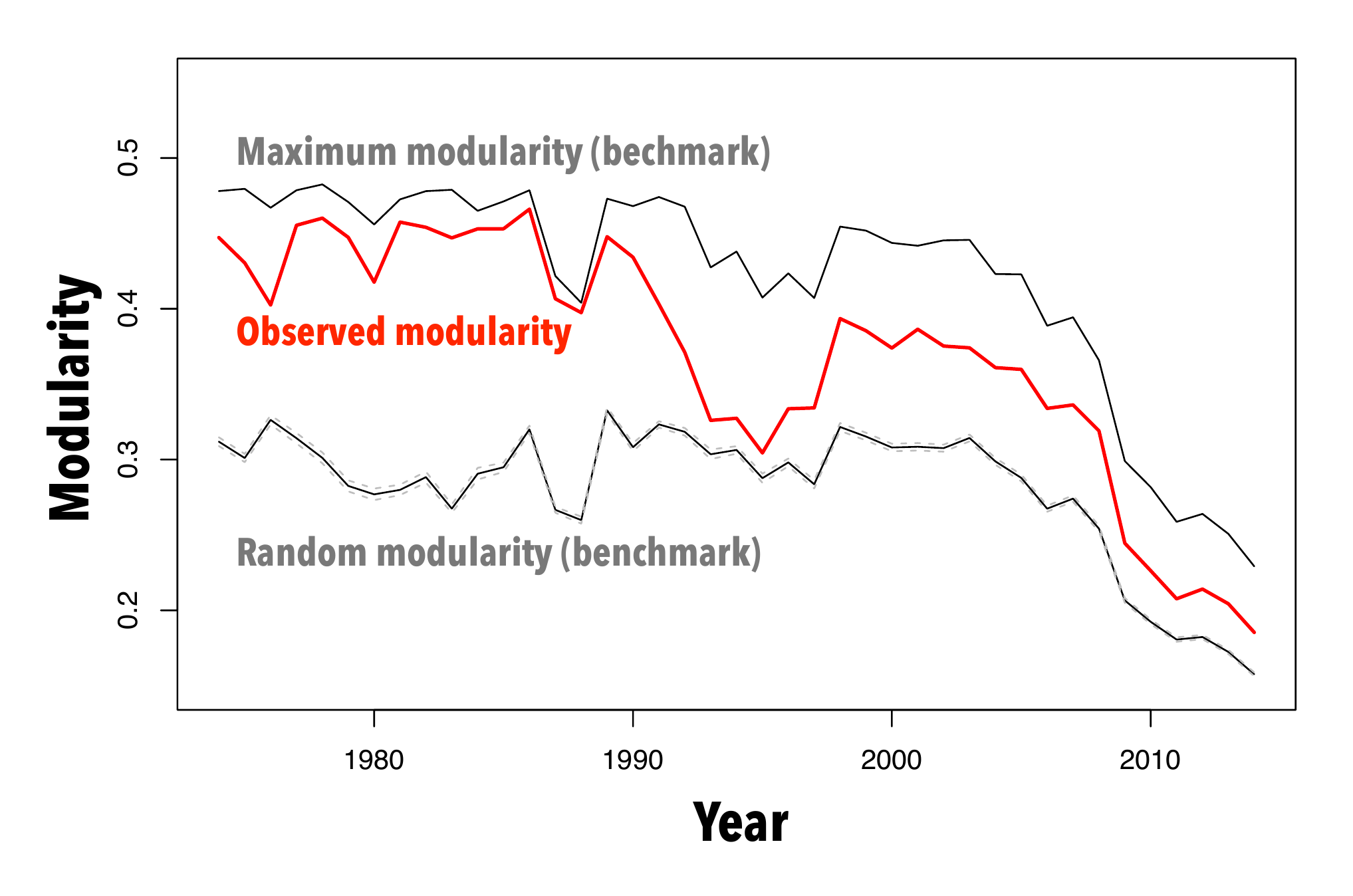}
  \caption{Maximum modularity, observed modularity (in red), and random modularity of the German research network. Random modularity (the modularity expected if connections were random) is shown with a 95\% confidence interval in grey dashed lines.}
  \label{Modularity}
\end{figure}

To study the modularity of the German research network we use equation~(\ref{Modularity}) with $s_{i}=+1$ for nodes belonging to West Germany and $s_{j}=-1$ for nodes belonging to East Germany. To establish a benchmark value of modularity, we created maximum and random modularity by allocating $s_{i}$ and $s_{j}$ to either maximize modularity, or at random. In both benchmarks we keep the same number of $s_{i} = +1$ and $s_{j} = -1$ than in the original network. To smooth out fluctuations, we generate 500 observations of random modularity for each year and average values and estimate a 95\% confidence interval. 

Figure~\ref{Modularity} shows the evolution of the modularity of the German research network in comparison with two benchmarks: maximum and random modularity. As shown in Figure~\ref{Modularity}, before reunification, the level of modularity was closer to the maximum benchmark, but it began decreasing after reunification in 1990. After 2000, however, modularity increased again and has been between the two benchmarks ever since. Even though the total modularity of the German research network, and the benchmarks, has been decreasing during the last 15 years. Figures ~\ref{Modularity} and ~\ref{dataSummary2}, tell us that the quick integration in the 1990s stagnated at the end of the twentieth century. 

To measure the speed of unification for each field, we define the speed as a percentage decrease in modularity, since lower value of modularity is corresponding to higher lever of integration. 
\begin{equation}
    \text{Speed}_{p} = - \frac{\mathcal{Q}_{t+\Delta t}-\mathcal{Q}_{t}}{\mathcal{Q}_{t}} \cdot 100
\label{speed}
\end{equation}
where $\text{Speed}_{p}$ is the speed of unification of field $p$, $\mathcal{Q}_{t}$ and $\mathcal{Q}_{t+\Delta t}$ are modularity at time $t$ and $t+\Delta t$, respectively.

\subsection{Measuring the balance and dominance of academic fields}

Next, we explore the factors that explain which fields achieved higher levels of reunification. We begin by measure the balance or dominance of each field prior to 1990. We do this by first identifying the regions which had revealed comparative advantage (RCA) in each field, following the method described in \cite{hidalgo2007product} and \cite{balassa1965trade}. The RCA of a region for a particular field is defined as the ratio between the observed number of publications for that region in that field and the number of publications expected, given the total academic output of that region and the total number of papers produced in that field. 

Formally, the RCA of region $r$ in field $p$ is (RCA is identical to what urban planners refer to as location quotient: LQ):
\begin{equation}
    RCA_{r,p} = \left.{\frac{x_{r,p}}{\sum_{p}x_{r,p}}}\middle/ \frac{\sum_{r}x_{r,p}}{\sum_{r,p}x_{r,p}}\right.
\label{RCA}
\end{equation}
where $x_{r,p}$ is the number publications of region $r$ in field $p$. Using this method, we estimated the RCA for each region and the 39 broad subject areas featured on the WoS.

We then selected the 20 fields with the largest number of publications before 1990. These fields are: clinical medicine; biological science; chemical science; physical science; basic medicine; materials engineering; other engineering and technologies; mathematics; health sciences; mechanical engineering; electrical, electronic, and information engineering; earth sciences and related environmental sciences; other natural sciences; veterinary science; agriculture, forestry, and fisheries; chemical engineering; computer and information sciences; medical engineering; environmental engineering; and animal and dairy science.

\linespread{1.5}
\begin{figure}[!t]
  \centering
  \includegraphics[width=1\textwidth]{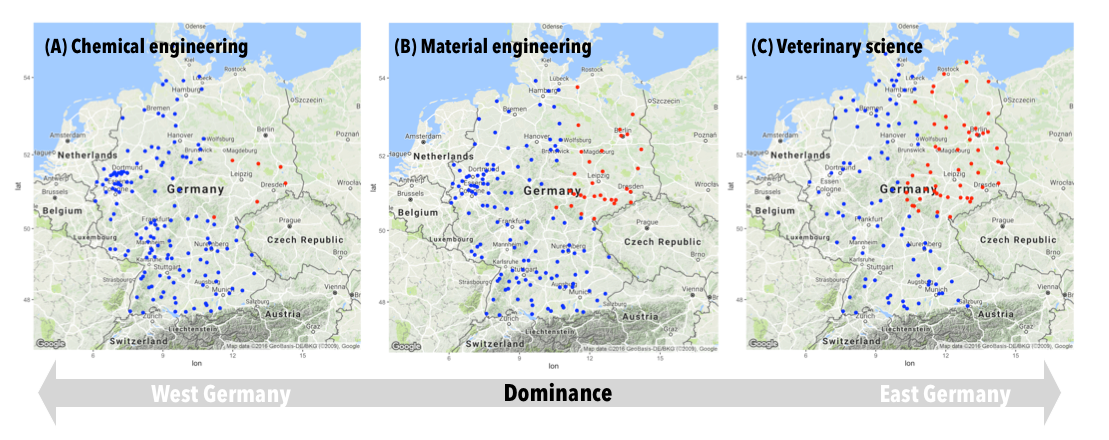}
  \caption{Regions that had Revealed Comparative Advantage (RCA $\geq$ 1) before 1990 in: (A) chemical engineering, (B) material engineering, and (C) veterinary science. Blue and red dots represent West and East German regions, respectively. Material engineering is a field that was dominated by West Germany prior to unification, while Veterinary Science was dominated by East Germany. }
  \label{RCAexample}
\end{figure}

Figure~\ref{RCAexample} shows the regions with a positive RCA in chemical engineering (A), material engineering (B), and veterinary science (C). Each dot in Figure~\ref{RCAexample} represents a region with RCA greater than or equal to 1, with regions in East Germany depicted by red dots and those in West Germany with blue dots. Material engineering is an example of a field dominated by West Germany prior to reunification, while Veterinary Science/ Chemical engineering is an example of a field dominated by East/West Germany.

Next, we estimated the dominance of each region for each academic subject by determining if, prior to 1990, most regions with a positive RCA in that field were located in East or West Germany. Formally, we defined the East or West dominance of a field as:
{\footnotesize
\begin{equation}
    Dominance_{p}= \frac{\textrm{\# regions in East with $RCA \geq 1$}}{71} -\frac{\textrm{\# regions in West with $RCA \geq 1$}}{270}
\label{Dominance}
\end{equation}
}
In equation~\ref{Dominance}, we adjust for the different sizes of East and West Germany by the number of regions in West (270) and East (71) Germany that had published at least one article in those fields before 1990.

\subsection{Connecting dominance and modularity}

\linespread{1.5}
\begin{figure}[!t]
  \centering
  \includegraphics[width=0.8\textwidth]{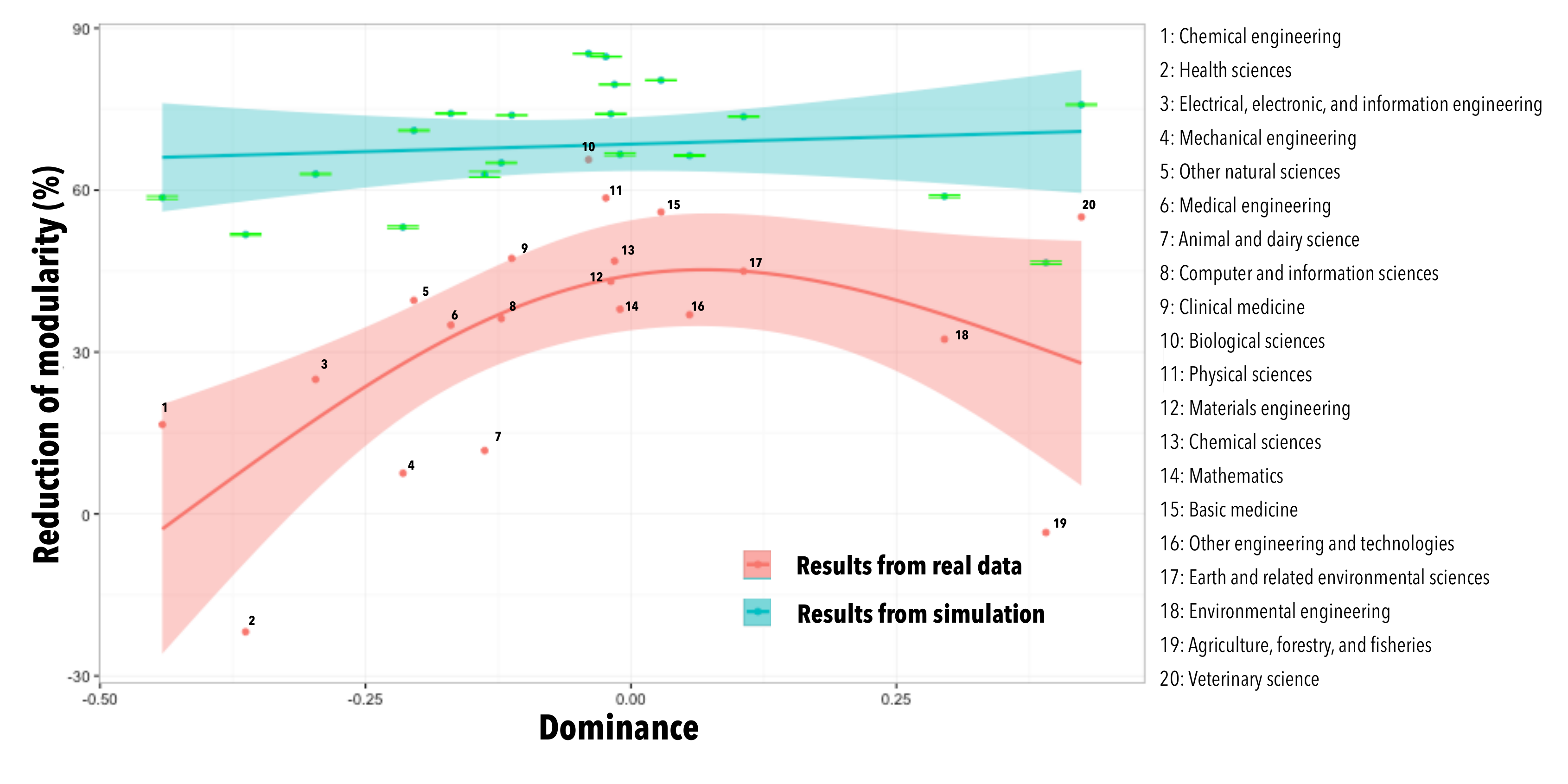}
  \caption{The relationship between the reduction in modularity of a field between 1985-2005 and the level of dominance of that field prior to reunification. Red dots present the values obtained from real data, while green dots represent those from the Monte Carlo simulations, with 95\% confidential intervals shown by green horizontal lines. Areas shaded in green and red represent the 95\% confidential intervals of the fits.}
  \label{mainResult}
\end{figure}

Next, we connect the change in modularity of a field with its dominance. Figure~\ref{mainResult} shows in red the relationship between the reduction in modularity in a field (y-axis) (a measure of subsequent integration) and its dominance prior to 1990 (x-axis). As a counterfactual, we present the values produced by the null model created through the Monte Carlo simulations (Green). In this Monte Carlo simulation we rewire links between regions for each field by preserving their total connectivity. This helps us compare the changes in modularity observed with those expected for the same regions if they had randomly connected to others.

\begin{figure}[!t]
  \centering
  \includegraphics[width=0.8\textwidth]{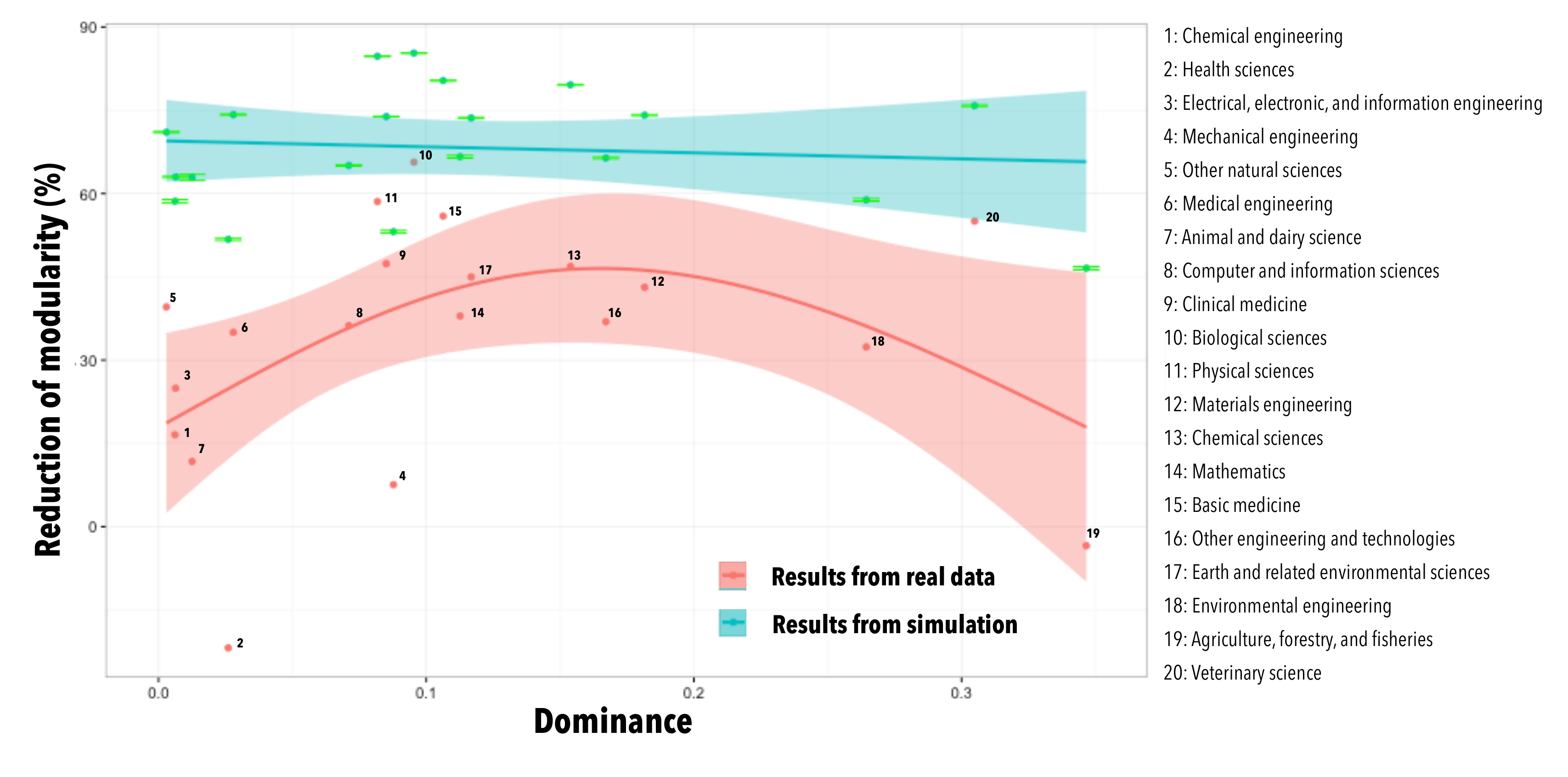}
  \caption{The relationship between the reduction in modularity in the period of 1985-2005 and the level of dominance of each academic field measured using  the share of publication East German publications in a field before 1990. Red dots present the values observed for the German research network while green dots present the counterfactual determined by a Monte Carlo simulation. Green error bars show the 95\% confidential intervals of the Monte Carlo simulation calculated using 500 iterations. Areas shaded in green and red represent the 95\% of confidential intervals around fitted lines.}
  \label{differentDominance}
\end{figure}

Comparing the results obtained from the original data with the Monte Carlo counterfactual reveals an inverted U-shaped relationship between the reduction in modularity of a field and the balance of that field prior to reunification. West- or East-skewed fields, such as chemical engineering, health science, agriculture, and veterinary science, experienced smaller changes in modularity after  reunification. More balanced fields, such as basic medicine, bio science, and material engineering, tended to reduce their modularity more. 

The Monte Carlo simulations help us test the significance and robustness of these results by checking whether the inverse U relationship should be expected simply from the variation in balance between the fields. For instance, one could expect that a field that is dominant in the West would simply have no regions to collaborate with in the East, and hence, could not experience a large change in its modularity. The Monte Carlo simulations tell us that the regions that were active in these fields, while preserving the number of connections they had, would have experienced much larger decreases in modularity than if they would have connected randomly to others. The lack of an inverse U relationship in the Monte Carlo simulation help us unsure that these reductions are characteristics of the evolution of the network of research collaborations, and not a relationship that should be expected from the definition of the variables or the network's starting conditions (see Appendix for details). Moreover, we check for robustness by repeating the exercise using the share of East Germany in the total number of publications in each field as a measure of dominance, obtaining a similar result (Figure \ref{differentDominance}). 

\subsection{Checking robustness using the differences-in-differences method}

Next, we check for the robustness of the inverted U-shaped by using a differences-in-differences (DID) regression. To perform a DID analysis we requires two groups: a treatment and a control group. In this paper, the treatment group includes all publications in the 10 most balanced fields, while the control group uses data on the 10 fields where the output was skewed to either East or West Germany.

\begin{figure}[!b]
  \centering
  \includegraphics[width=0.7\textwidth]{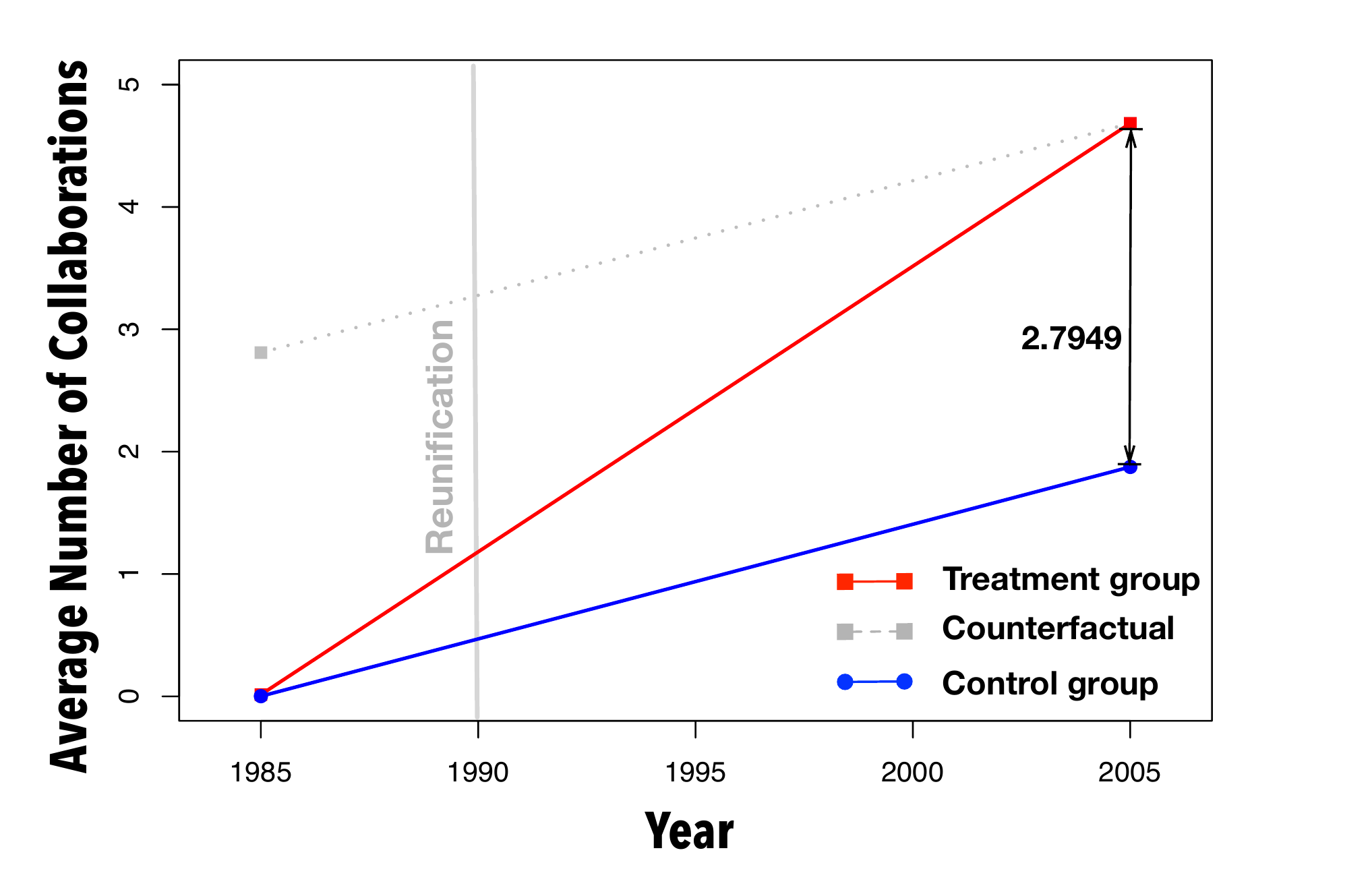}
  \caption{Average number of collaborations in 1985 and 2005. The treatment group is represented by the red line, while the control group is shown in blue. As the average number of collaborations in both the treatment and control group is almost zero, and the counterfactual line moves in parallel with that of the control group.}
  \label{DIDresult}
\end{figure}

One might worry about the possibility of selection bias in establishing the two groups, since DID is most appropriately used when an event is close to random and features conditional time and group-fixed effects \citep{Bertrand2004}. In our case, the level of dominance that determines the group of fields is calculated without considering collaboration networks and by focusing only on the number of papers produced in each region and in each field before 1990, while the increase in the number of collaborations between East and West after 1990 is the dependent variable in our DID-based empirical model. Therefore, the manner in which we calculate the level of dominance does not directly guarantee faster collaboration after reunification. Moreover, as seen in Appendix C, other variables, such as population density, per capita GDP, and distances between regions, are similar between the two groups. 

To check the different effects of reunification, we selected two years, 1985 and 2005: one before and one after reunification. Basically, we used the same data that we used in previous sections, but we only considered the West-East collaboration, ignoring the collaborations within the West and within the East. However, we were unable to obtain data for other control variables, such as population density and per capita GDP before the reunification, even for the NUTS 2 regions. Consequently, to control for other factors we analyzed the different effects of reunification on the number of collaborations between 1995 and 2005. Furthermore, we controlled for the geographic/geodesic distance between the two regions using the regions' coordinates. Summary statistics in this regard are provided in Table~\ref{tab:DID_apendix1} and \ref{tab:DID_apendix2}.

Then, is the inverted U-shaped relationship statistically significant? In this section, we examine the robustness of the finding that fields with balanced capabilities achieved faster unification in their collaboration networks after 1990 compared to the fields with skewed research capabilities using a differences-in-differences (DID) method. This method can verify that the effect of reunification in 1990 on the collaboration networks is statistically different between the fields with balanced dominance and those with skewed dominance. Because we look at not the change in modularity but the change in real number of collaboration, this robustness check ensures that the inverted U-shaped relationship is not driven by the usage of modularity. 

DID requires two groups: a treatment and a control group. In this paper, the treatment group includes all publications in 10 fields of which the level of capabilities of the West and East are balanced, as we have already presented in Figure~\ref{mainResult}. Those fields are located around zero of the x-axis in Figure~\ref{mainResult}. On the other hand, our control group covers the other 10 fields of study that represent West- or East-skewed fields in terms of their dominance in Figure~\ref{mainResult}. 

The DID method is widely used to estimate the causal effects of events or policy interventions. However, the aim of this study is not to measure the effect of reunification, but to compare the different levels of effects to the speed of unification in terms of the capability balance of fields. Therefore, our focus is not on checking the cause of the event, which is the reunification of 1990, but to verify that the effect is significantly different for the two groups: treatment and control.

One might worry about the possibility of selection bias in establishing the two groups, since DID is appropriate to use when an event is close to random and conditional on time and group-fixed effects \citep{Bertrand2004}. In our case, the level of dominance that determines the group of fields is calculated without considering collaboration networks and by focusing only on the number of papers in each region and in each field before 1990, while the increase in the number of collaborations between East and West after 1990 is our dependent variable in our empirical model using DID. Therefore, the way that we calculate the level of dominance does not directly guarantee faster collaboration after reunification. Moreover, as seen in Appendix C, the other variables, such as population density, per capita GDP, and distance between the regions, are similar between the two groups. 

To check the different effects of reunification, we selected two years, 1985 and 2005: one is before and the other is after reunification. Basically, we used the same data that we used in previous sections, but we only considered the West-East collaboration, ignoring the collaborations within West and within East. However, we could not find data for other control variables, such as population density and per capita GDP before the reunification, even in the NUTS 2 region. To control for other factors, therefore, we analyzed the different effects of reunification on the number of collaborations for the years 1995 and 2005. Also, we controlled for the geographic/geodesic distance between the two regions using the regions' coordinates. Table~\ref{tab:DID_apendix1} and \ref{tab:DID_apendix2} provides summary statistics. 	

\begin{table}[!t]
  \centering
  \caption{The result of differences-in- differences analysis (for 1985 and 2005)}
     \footnotesize
    \begin{tabular}{lccccccc}
          &       &       &       &       &       &       &  \\
    \midrule
    \midrule
    \multirow{3}[6]{*}{} & \multicolumn{2}{c}{DID} & \multicolumn{5}{c}{Add covariates} \\
\cmidrule{2-8}          & 1985-2005 & 1995-2005 & 1985-2005 & \multicolumn{4}{c}{1995-2005} \\
\cmidrule{2-8}          & (1)   & (2)   & (3)   & (4)   & (5)   & (6)   &  (7)\\
    \midrule
    \multirow{2}[1]{*}{Interaction} & 2.7949*** & 2.1504*** & 2.7949*** & 2.1504*** & 2.4105*** & 2.1458*** & 2.4057*** \\
          & (0.2275) & (0.2370) & (0.2276) & (0.2372) & (0.3211) & (0.2374) & (0.3218) \\
    \multirow{2}[0]{*}{Treatment} & 1.1695*** & 1.8589*** & 1.0281*** & 1.6826*** & 2.2747*** & 1.7963*** & 1.8798*** \\
          & (0.2129) & (0.2302) & (0.2088) & (0.2240) & (0.3310) & (0.2296) & (0.3199) \\
    \multirow{2}[0]{*}{After} & 1.8739*** & 1.7541*** & 1.8739*** & 1.7541*** & 1.8162*** & 1.7611*** & 1.8108*** \\
          & (0.0837) & (0.0856) & (0.0841 & (0.0861) & (0.1089) & (0.0881) & (0.1150) \\
    \multirow{2}[0]{*}{Distance} &       & \multirow{2}[0]{*}{} & 1.2767** & 1.5918** &       &       & 3.0966*** \\
          &       &       & (0.5861) & (0.6137) &       &       & (0.9619) \\
    $\Delta$population &       & \multirow{2}[0]{*}{} &       &       & -0.1124 &       & -0.4192*** \\
    (log scale) &       &       &       &       & (0.1182) &       & (0.1427) \\
    $\Delta$GDP  &       & \multirow{2}[0]{*}{} &       &       &       & 0.4816*** & 0.5579*** \\
    (log scale) &       &       &       &       &       & (0.0917) & (0.1333) \\
     \midrule
    Observations & 8110  & 8110  & 8110  & 8110  & 5717  & 8110  & 5717 \\
    Robust R2 & 0.0864 & 0.0690 & 0.0868 & 0.0696 & 0.0656 & 0.0724 & 0.0704 \\
    RMSE  & 7.7172 & 8.0612 & 7.7159 & 8.0591 & 9.1516 & 8.047 & 9.1295 \\
    \bottomrule
    \bottomrule
    \end{tabular}%
    \vspace{1ex}
     \raggedright We use the DID method to conduct this analysis. The variable $interaction$ captures the difference in the impact of German reunification between the treatment and control groups (regional pairs from the balanced fields and from the skewed fields in terms of dominance, respectively). $Treatment$ is a dummy variable indicating whether each regional pair originates from the balanced fields, while $After$ is a dummy variable indicating whether information on regional pairs originates from 2005. Population and GDP of each region are in a natural log scale. Standard errors are clustered at the regional-pair level. $*p < 0.10$, $**p<0.05$, $***p <0.01$.
  \label{DIDResultTable}
\end{table}

The empirical specification for the ordinary least squares (OLS) estimation using the DID method is provided below, with $y_{ij}$ denoting the number of collaborations between regions $i$ and $j$, and $\mathbf{X}$ including other control variables. Again, because we only consider collaborations between East and West, $i$ and $j$ refer to region $i$ in West Germany and region $j$ in East Germany, respectively.
\begin{equation}
y_{ij}=\alpha+\beta(Treat_{ij}*After_{ij})+\sigma Treat_{ij}+\lambda After_{ij} +\mathbf{B}\mathbf{X}^{\prime}+\varepsilon_{i,j}
\end{equation}

In this empirical model, the positive sign of the coefficient of the interaction term, $\beta$, means that the probability of further collaboration increases for more balanced fields. The coefficients $\sigma$ and $\lambda$ explain the differences in the number of collaborations between the two groups and between the two time periods, with one before and one after reunification. 

Figure~\ref{DIDresult} and Table~\ref{DIDResultTable} show the result of the DID analysis. Figure~\ref{DIDresult} informs us that there were few collaborations between East and West in 1985 but that this number increased after reunification. The most balanced fields increased their number of East-West collaborations between two to three more (2.14 to 2.79 times more) than the less balanced fields.  This means that, in the fields whose output was more balanced between  East and West, scientists were more likely to establish collaborations after reunification. Table~\ref{DIDResultTable} shows that this difference is statistically significant. The result holds even when controlling for demographic and economic factors (which are available starting in 1995), such as the differences in GDP and population density, as well as geographic distance, which is the shortest distance between the two regions, as seen in columns (3) through (7). 

\subsection{Why do they meet in the middle?}
\linespread{1.5}
\begin{figure}[ht]
  \centering
  \includegraphics[width=0.7\textwidth]{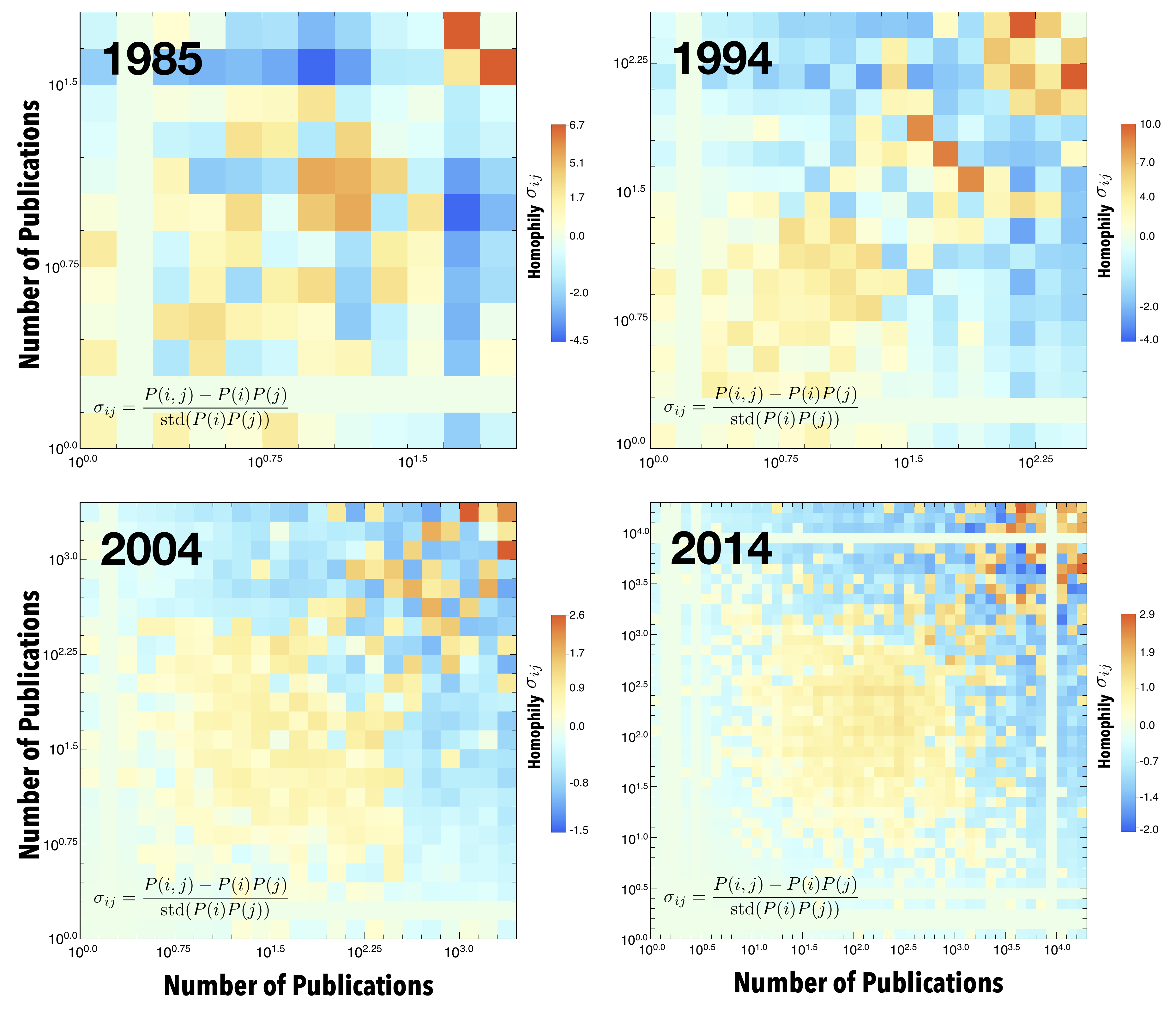}
  \caption{Mixing pattern of networks in 1985, 1994, 2004, and 2014. The color identifies the correlation between regions with different scientific productivity, i.e., the number of publications.}
  \label{MixingPattern}
\end{figure}

Finally, we explore one mechanism that could help explaining why the more balanced fields integrated more effectively than those with a more skewed output. This is the homophily principle: the idea that "contact between similar people occurs at a higher rate than among dissimilar people," \citep{McPherson2003}. The homophily principle has been shown to operate in various types of relationships, such as marriage \citep{Kalmijn2003}, friendship \citep{Verbrugge1977}, career support \citep{Ibarra1992, Ibarra1995}, networking in social media \citep{DeChoudhury2010}, and even mutual attraction in a public place \citep{Mayhew1995}. 

Since we cannot trace individuals scholars, we test the role of homophily by studying the mixing patterns of NUTS3 regions. That is, we explore whether regions were more or less likely to establish collaborations with other regions with the same levels of productivity (as measured by the fraction of papers they produced in a field). 

\begin{figure}[ht]
  \centering
  \includegraphics[width=1\textwidth]{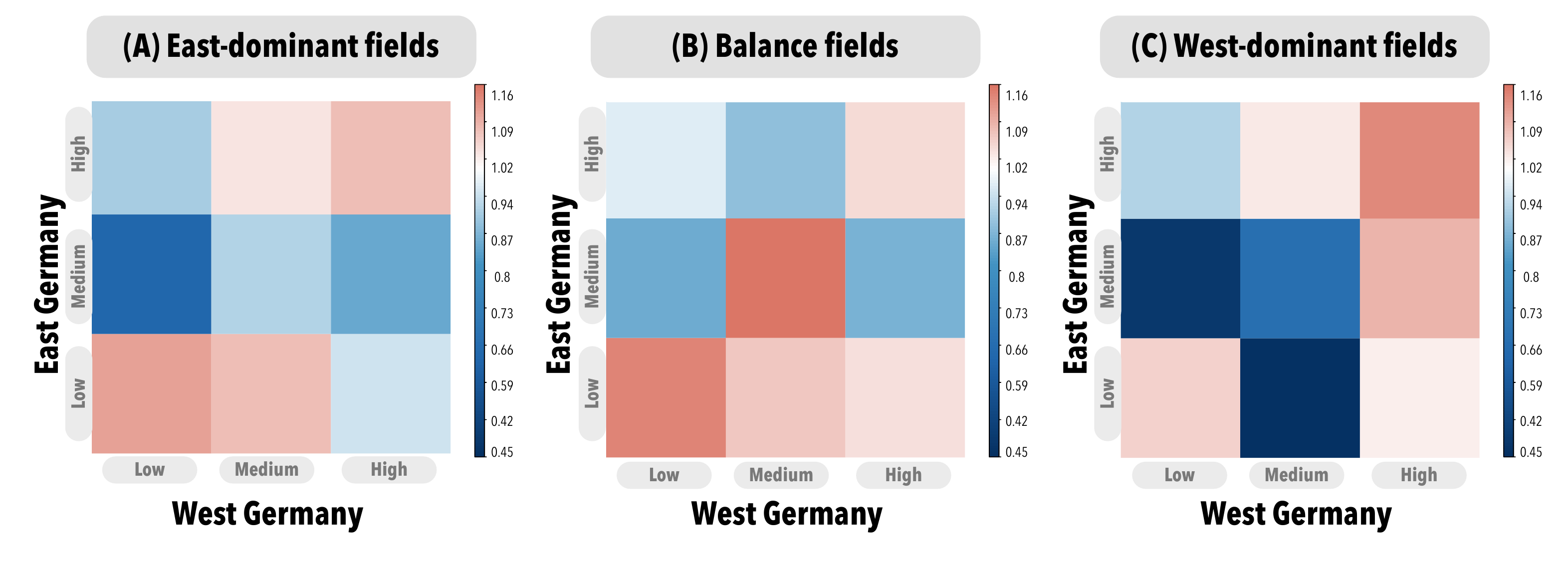}
  \caption{Collaboration patterns of (A) East-dominant fields, (B) balanced fields, and (C) West-dominant fields. We use a diverging color scale with white indicating as many collaborations as expected by chance, red more collaborations than expected by chance, and blue, for when we observe less collaborations than expected by chance.}
  \label{whyInvertedU}
\end{figure}

Figure~\ref{MixingPattern} shows the mixing patterns of networks with respect to the scientific productivity of regions during the years of 1985, 1994, 2004, and 2014. Following \cite{Maslov2005}, the matrices show a measure of Homophily ($\sigma$) comparing the number of co-authorships between regions after controlling for the expected number of co-authorships, and binned by their number of publications. 
\begin{equation}
\sigma_{i,j}=\frac{P(i,j)-P(i)P(j)}{std(P(i)P(j))}
\label{MixingPattern_all}
\end{equation}
The fact that most positive values (yellow, orange, and red), concentrate along the diagonal (Figure~\ref{MixingPattern}) tells us that regions with similar levels of productivity were more likely to collaborate than what we would expect by random chance. 

Next, we check this homophily characteristic by focusing only on East-West collaborations. This time we calculate the RCA of each region in a field, so we can look at the number of collaborations among regions with similar, or different, levels of RCA. For each field we separate regions into three categories, "low," "medium," and "high," depending on their RCA value using quartiles ("low" for those below the first quartile, "high" for those above the third quartile, and "medium" for those within the second and third quartile). We then count the number of collaborations among these three groups, and determine the number of collaborations among each pair of RCA categories (low-low, low-middle, etc.). Again, we normalize the total number of observed collaborations with its expected number, obtained from the average of 100 Monte Carlo simulations. 

Figure~\ref{whyInvertedU} shows whether each region had more or less collaborations than expected by chance between "low," "medium," and "high" comparative advantage regions. In the case of balanced fields (Figure~\ref{whyInvertedU} B), we observe a strong tendency for regions to collaborate with other regions with a similar level of comparative advantage. Once again, this provides evidence of homophily in the research network, suggesting that a researcher in a balanced field might be able to find a similar researcher in terms of research capacity from another part of Germany, which could lead to further collaboration between West and East Germany.

\section{Conclusion}

In this paper, we explored the evolution of co-authorship network among regions in East and West Germany between 1974 and 2014. We find that the integration of the German research network, after 1990, was initially fast, but then stagnated at an intermediate level of integration. Then we explored the factors affecting the level of integration of a field by studying the reduction in modularity of each field and comparing it with the balance of its output prior to reunification. We found that the reduction of modularity in a field followed an inverse U-relationship as a function of a field's dominance. In other words, fields dominated by either West or East Germany prior to reunification integrated less than fields that possessed a more balanced output. This was confirmed by using Monte Carlo simulations and a DID analysis. 

Finally, we explored whether homophily could have contributed to this inverse U relationship between a field's level of integration and its balance prior to 1990. We found an assortative mixing pattern, meaning that regions are more likely to collaborate with regions with a similar productivity, even after controlling for the overall productivity of each region in each field. This suggests that after 1990, it may have been easier for researchers from the balanced fields to find collaborators from the other side of Germany, while those in the skewed fields might have experienced difficulty, leading to a slower speed of integration. 

While encouraging, our results should be interpreted in light of their limitations. For instance, we aggregate the data into the NUTS 3 regional level, even though our data include additional author information, such as the names of institutions the authors were affiliated with. As such, this aggregation prevents us from examining various reunification channels relating to the German research network; i.e., we do not provide a micro-level pattern of research collaboration. For example, we cannot capture instances where a researcher in West Germany moved to East Germany after reunification and held an Eastern German address. We believe this to be an important limitation since East-West migration was large after, and hence, many of the collaboration links could also be a consequence of migration. Our results do not contradict the migration hypothesis, but are not able to provide evidence of it because of the resolution of our data (it is not disambiguated at the author level). 

Additionally, our data is limited. As the data only captures publications from the journals listed on the Science Citation Index of WoS, other forms of research collaboration, especially those in the German language, are missing. Considering that Germany has a strong research tradition in its own language, we cannot say that our study encompasses all major research collaboration.

Nevertheless, our findings on the evolution of research collaboration help to expand the body of literature on research collaboration, knowledge diffusion, and the unification of a research system when two systems are merged. Moreover, unveiling the factors that affect the speed of unification could have beneficial policy implications for the countries that are still awaiting political unification, such as the two Koreas.

\section*{Acknowledgments}

We wish to acknowledge the support received from the MIT Media Lab Consortia and from the Masdar Institute of Technology. We would also like to thank Xiaowen Dong, Mary Kaltenberg, Aamena Alshamsi, Jian Gao, and Juan Carlos Rocha for their valuable comments. This work was supported by Korea Institute of S\&T Evaluation and Planning (KISTEP), the Center for Complex Engineering Systems (CCES) at King Abdulaziz City for Science and Technology (KACST), and the Massachusetts Institute of Technology (MIT).

\clearpage
\linespread{1.5}
\bibliographystyle{elsarticle-harv}
\biboptions{authoryear}
\bibliography{reference}

\begin{thebibliography}{36}
\expandafter\ifx\csname natexlab\endcsname\relax\def\natexlab#1{#1}\fi
\expandafter\ifx\csname url\endcsname\relax
  \def\url#1{\texttt{#1}}\fi
\expandafter\ifx\csname urlprefix\endcsname\relax\def\urlprefix{URL }\fi

\bibitem[{Ardagh(1988)}]{ardagh1988germany}
Ardagh, J., 1988. {Germany and the Germans: an anatomy of society today}.
  Harper Perennial.

\bibitem[{Bach and Trabold(2000)}]{Bach2000}
Bach, S., Trabold, H., 2000. {Ten years after German monetary, economic and
  social union: an introduction }. Quarterly Journal of Economic Research
  69~(2), 149--151.

\bibitem[{Balassa(1965)}]{balassa1965trade}
Balassa, B., 1965. Trade liberalisation and “revealed” comparative
  advantage. The Manchester School 33~(2), 99--123.

\bibitem[{Bertrand et~al.(2004)Bertrand, Duflo, and
  Mullainathan}]{Bertrand2004}
Bertrand, M., Duflo, E., Mullainathan, S., Feb. 2004. {How Much Should We Trust
  Differences-In-Differences Estimates?} The Quarterly Journal of Economics
  119~(1), 249--275.

\bibitem[{BMBF(2015)}]{BMBF}
BMBF, 2015. {Programm-Monitoring, DLR Projektträger, Stand November 2015 in
  “25 Jahre Deutsche Einheit – 15 Jahre Unternehmen Region”}. 7th KISTEP
  Future Forum.

\bibitem[{BMFT(1990)}]{BMFT}
BMFT, 1990. {Pressemitteilung}.

\bibitem[{BMWi(2015)}]{BMWi2015}
BMWi, 2015. {Wirtschaftsdaten neue bundesländer}. Available at
  https://www.bmwi.de/BMWi/Redaktion/PDF/W/wf-wirtschaftsdaten-neue-
  laender,property=pdf,bereich=bmwi2012,sprache=de,rwb=true.pdf.

\bibitem[{Borgatti et~al.(2009)Borgatti, Mehra, Brass, and
  Labianca}]{Borgatti2009}
Borgatti, S.~P., Mehra, A., Brass, D.~J., Labianca, G., Feb. 2009. {Network
  Analysis in the Social Sciences}. Science 323~(5916), 892--895.

\bibitem[{Christakis and Fowler(2009)}]{christakis2009connected}
Christakis, N.~A., Fowler, J.~H., 2009. Connected: The surprising power of our
  social networks and how they shape our lives. Little, Brown.

\bibitem[{De~Choudhury et~al.(2010)De~Choudhury, Sundaram, and
  John}]{DeChoudhury2010}
De~Choudhury, M., Sundaram, H., John, A., 2010. {" Birds of a Feather": Does
  User Homophily Impact Information Diffusion in Social Media?} arXiv.org.

\bibitem[{Eickelpasch(2015)}]{eickelpasch2015forschung}
Eickelpasch, A., 2015. Forschung, entwicklung und innovationen in
  ostdeutschland: R{\"u}ckstand strukturell bedingt. DIW-Wochenbericht 82~(41),
  907--918.

\bibitem[{Fortunato(2010)}]{Fortunato2010}
Fortunato, S., Feb. 2010. {Community detection in graphs}. Physics Reports
  486~(3-5), 75--174.

\bibitem[{Gareau(1961)}]{gareau1961morgenthau}
Gareau, F.~H., 1961. Morgenthau's plan for industrial disarmament in germany.
  Western Political Quarterly 14~(2), 517--534.

\bibitem[{Granovetter(1973)}]{granovetter1973strength}
Granovetter, M.~S., 1973. The strength of weak ties. American journal of
  sociology, 1360--1380.

\bibitem[{G{\"u}nther et~al.(2010)G{\"u}nther, Nulsch, and
  Urban-Thielicke}]{Gunther2010}
G{\"u}nther, J., Nulsch, N., Urban-Thielicke, D., 2010. {20 Jahre nach dem
  Mauerfall: Transformation und Erneuerung des ostdeutschen
  Innovationssystems}. Studien zum deutschen {\ldots}.

\bibitem[{Hidalgo(2016)}]{Hidalgo2016}
Hidalgo, C.~A., Jul. 2016. {Disconnected, fragmented, or united? a
  trans-disciplinary review of network science}. Applied Network Science 1~(1),
  6.

\bibitem[{Hidalgo et~al.(2007)Hidalgo, Klinger, Barab{\'a}si, and
  Hausmann}]{hidalgo2007product}
Hidalgo, C.~A., Klinger, B., Barab{\'a}si, A.-L., Hausmann, R., 2007. The
  product space conditions the development of nations. Science 317~(5837), 482.

\bibitem[{Ibarra(1992)}]{Ibarra1992}
Ibarra, H., Sep. 1992. {Homophily and Differential Returns: Sex Differences in
  Network Structure and Access in an Advertising Firm}. Administrative Science
  Quarterly 37~(3), 422.

\bibitem[{Ibarra(1995)}]{Ibarra1995}
Ibarra, H., Jun. 1995. {Race, Opportunity, And Diversity Of Social Circles In
  Managerial Networks}. Academy of management journal 38~(3), 673--703.

\bibitem[{Kalmijn(2003)}]{Kalmijn2003}
Kalmijn, M., Nov. 2003. {Intermarriage and Homogamy: Causes, Patterns, Trends}.
  dx.doi.org 24~(1), 395--421.

\bibitem[{Lancichinetti and Fortunato(2011)}]{Fortunato2011}
Lancichinetti, A., Fortunato, S., Dec. 2011. {Limits of modularity maximization
  in community detection}. Phys. Rev. E 84~(6), 066122.

\bibitem[{Maslov and Sneppen(2005)}]{Maslov2005}
Maslov, S., Sneppen, K., Dec. 2005. {Computational architecture of the yeast
  regulatory network}. Physical biology 2~(4), S94--S100.

\bibitem[{Mayhew et~al.(1995)Mayhew, McPherson, Rotolo, and
  Smith-Lovin}]{Mayhew1995}
Mayhew, B.~H., McPherson, M., Rotolo, T., Smith-Lovin, L., 1995. {Sex and
  ethnic heterogeneity in face-to-face groups in public places: an ecological
  perspective on social interaction}. Social Forces.

\bibitem[{McPherson et~al.(2003)McPherson, Smith-Lovin, and
  Cook}]{McPherson2003}
McPherson, M., Smith-Lovin, L., Cook, J.~M., Nov. 2003. {Birds of a Feather:
  Homophily in Social Networks}. Annual Review of Sociology 27~(1), 415--444.

\bibitem[{Meske(1993)}]{Meske1993}
Meske, W., Oct. 1993. {The restructuring of the East German research system
  {\textemdash} a provisional appraisal}. Science and Public Policy 20~(5),
  298--312.

\bibitem[{Newman(2003)}]{Newman2003}
Newman, M., Jan. 2003. {The Structure and Function of Complex Networks}. SIAM
  Rev. 45~(2), 167--256.

\bibitem[{Newman(2001{\natexlab{a}})}]{Newman2001}
Newman, M.~E., 2001{\natexlab{a}}. {The structure of scientific collaboration
  networks}. Proceedings of the National Academy of Sciences 98~(2), 404--409.

\bibitem[{Newman(2006)}]{Newman2006}
Newman, M.~E., 2006. {Finding community structure in networks using the
  eigenvectors of matrices}. Physical review E 74~(3), 036104.

\bibitem[{Newman(2001{\natexlab{b}})}]{Newman12001}
Newman, M. E.~J., Jun. 2001{\natexlab{b}}. {Scientific collaboration networks.
  I. Network construction and fundamental results}. Phys. Rev. E 64~(1),
  016131.

\bibitem[{Newman(2001{\natexlab{c}})}]{Newman22001}
Newman, M. E.~J., Jun. 2001{\natexlab{c}}. {Scientific collaboration networks.
  II. Shortest paths, weighted networks, and centrality}. Phys. Rev. E 64~(1),
  016132.

\bibitem[{Pence and Betts(2008)}]{pence2008socialist}
Pence, K., Betts, P., 2008. Socialist modern: East German everyday culture and
  politics. University of Michigan Press.

\bibitem[{Rainie and Wellman(2012)}]{rainie2012networked}
Rainie, L., Wellman, B., 2012. Networked: The new social operating system. Mit
  Press.

\bibitem[{Storper and Venables(2004)}]{Storper2004}
Storper, M., Venables, A.~J., 2004. Buzz: Face-to-face contact and the urban
  economy. Journal of Economic Geography 4~(4), 351--370.

\bibitem[{Verbrugge(1977)}]{Verbrugge1977}
Verbrugge, L.~M., Dec. 1977. {The Structure of Adult Friendship Choices}.
  Social forces 56~(2), 576.

\bibitem[{Weber(2002)}]{weber2002kleine}
Weber, J., 2002. Kleine Geschichte Deutschlands seit 1945. Vol. 30830.
  Deutscher Taschenbuch Verlag.

\bibitem[{Wrobel(2010)}]{Wrobel2010}
Wrobel, R.~M., 2010. The benefits of german unification: A review after 20
  years. Seminar “20 years of German unification and Korea’s unification
  readiness”.

\end{thebibliography}

\clearpage

\setcounter{figure}{0}
\setcounter{table}{0}
\setcounter{equation}{0}
\setcounter{page}{1}
\setcounter{section}{0}
\renewcommand{\thefigure}{A\arabic{figure}}
\renewcommand{\thetable}{A\arabic{table}}
\renewcommand\theequation{A\arabic{equation}}






\section*{Appendix A. Network representation of research collaboration between West and East Germany}

We use the publication data to build a network where nodes are German regions (at the NUTS3 level; Eurostat 2003) and links connect pairs of regions when the scientists of these regions have published a paper together. For the network representation in Figure~\ref{dataSummary1} and the calculation of modularity without considering different fields of study in Figure~\ref{Modularity}, we focus only on links that are statistically significant, since the links between German regions vary widely in the number of co-authorships involved in them. When we examine the 20 fields of study, we consider all the links because of smaller number of observations. 

To identify significant links, we use filtering techniques from the network science literature that are based on comparing the observed strength of a link with its expected value. This allows us to separate the links that are overexpressed--and represent strong connections between regions--from those that are underexpressed, and could be explained by chance. 
\begin{figure}[!b]
  \centering
  \includegraphics[width=1\textwidth]{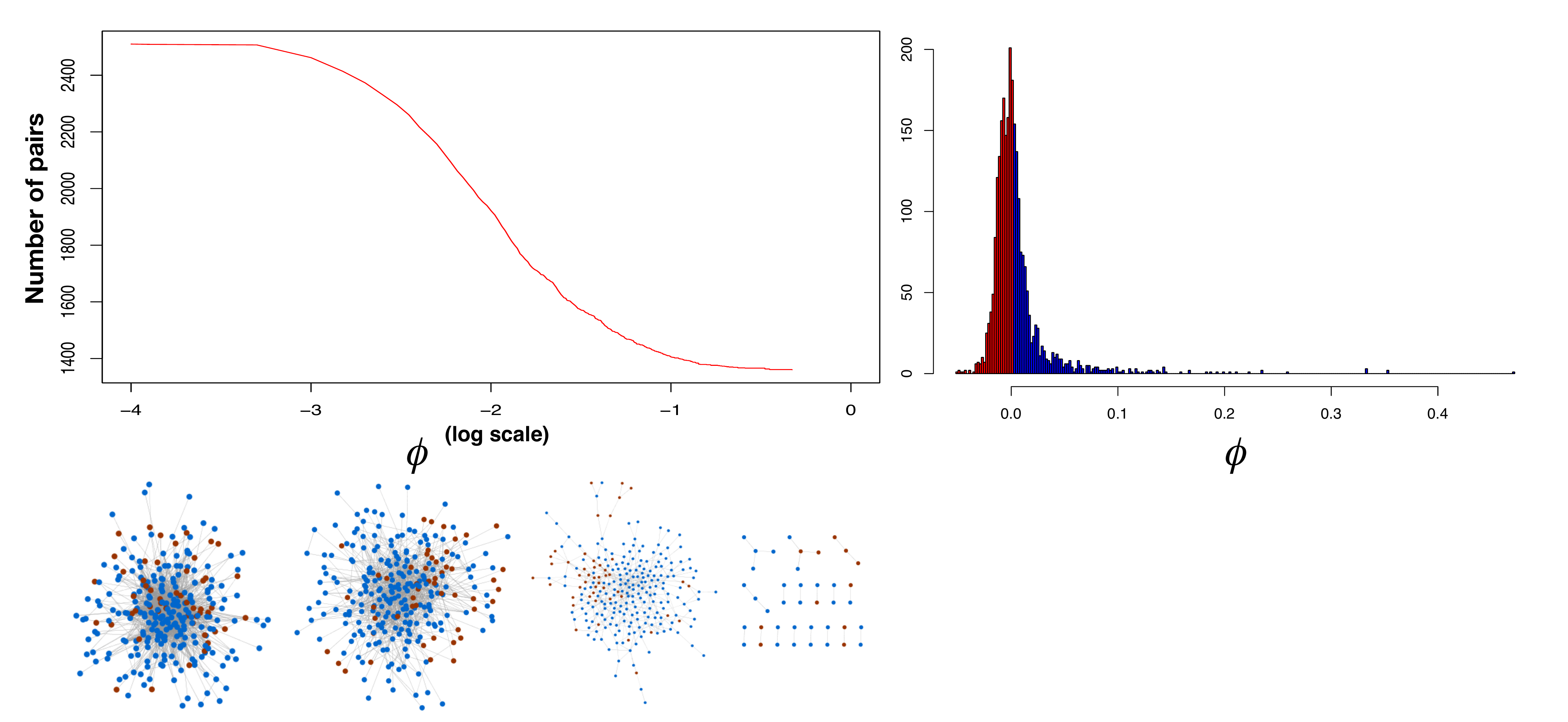}
  \caption{(Left) Number of regional pairs in the largest component of the network (in 1994) as a function of the  threshold. The lower panel shows 4 examples of networks at different value of $\phi$, which are $\phi=10^{-4}, 10^{-3}, 10{^-2}$, and $10^{-1}$, respectively. Nodes in red represent East German regions, while nodes in blue represent West German regions. (Right) Frequency of $\phi$ value}
  \label{phiValue}
\end{figure}
We estimate the strength of each link by calculating the  $\phi$-correlation coefficient associated with a pair of regions, which is Pearson's correlation for binary variables. The $\phi$-correlation can be expressed mathematically as:
\begin{equation}
 \phi_{ij} = \frac{C_{ij} N - N_i N_j}{\sqrt{N_i N_j \left( N - N_i \right) \left( N - N_j \right)}}
\end{equation}
where $C_{ij}$is the number of links between region $i$ and $j$; $N$ is the total number of authors who publish in that year; and $N_{i}$ and $N_{j}$ is the number of authors in region $i$ and $j$, respectively. 

We can determine the significance of $\phi \neq0$ using a $t$-test, where $t$ is: 
\begin{equation}
 t =\frac{\phi_{ij} \sqrt{n - 2}}{\sqrt{1 - \phi_{ij}^2}}
\end{equation}
where $n$ is the number of observations used to calculate $\phi$. In this study, $n$ is the total number of authors in each year. As a rule of thumb, when $n$ is greater than 1,000, any link with   is significant at the 5\% level, while those with  are significant at the 1\% level. So we focus only on links with $t$ greater or equal to 2.58. 

The distribution and frequency of $\phi$ values representing all links where $C_{ij} > 0$ are presented in Figure~\ref{phiValue}. 	

\setcounter{figure}{0}
\setcounter{table}{0}
\setcounter{equation}{0}
\renewcommand{\thefigure}{B\arabic{figure}}
\renewcommand{\thetable}{B\arabic{table}}
\renewcommand\theequation{B\arabic{equation}}

\clearpage
\section*{Appendix B. Rewiring network by the Monte Carlo Simulation}

\begin{figure}[h]
  \centering
  \includegraphics[width=0.75\textwidth]{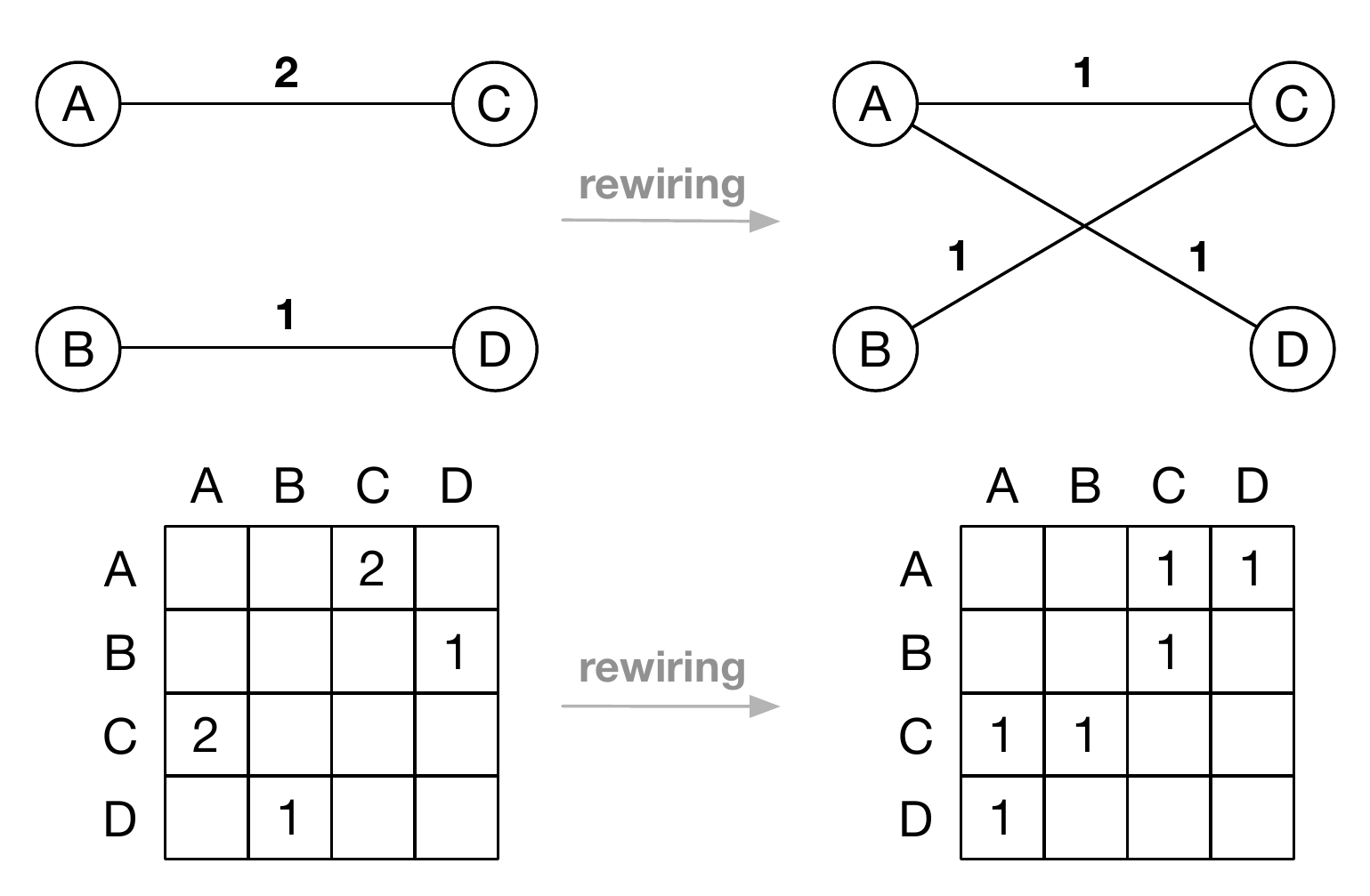}
  \caption{Rewiring network with keeping each node's degree unchanged}
  \label{rewiring}
\end{figure}

To have a benchmark network in evolution of collaboration network, we build a null model by randomization of network with the Monte Carlo simulation. In every iteration, we select four nodes by random sampling, and rewire its links as described in Figure~\ref{rewiring} with keeping the degree same. This means that we randomize not the number of collaboration of each region or regions' research capabilities, but selection of partners for collaboration. For generating one rewired observation, we randomize the network with 1,000 times of edge list's length as a number of iteration. We create hundred observations, and consider its mean and standard errors.

\setcounter{figure}{0}
\setcounter{table}{0}
\setcounter{equation}{0}
\renewcommand{\thefigure}{C\arabic{figure}}
\renewcommand{\thetable}{C\arabic{table}}
\renewcommand\theequation{C\arabic{equation}}

\clearpage
\begin{landscape}
\section*{Appendix C. Differences-in-differences analysis}
\begin{table}[htbp]
  \centering
  \caption{Summary statistics of entire links, including within West, within East and between East and West (1995-2005)}
    \begin{tabular}{lccccc}
          &       &       &       &       &  \\
    \midrule
    \midrule
    \multicolumn{1}{c}{\multirow{2}[3]{*}{Entire links}} & \multicolumn{2}{c}{Before} & \multicolumn{2}{c}{After} & \multirow{2}[3]{*}{Differences in Differences} \\
\cmidrule{2-5}          & Control group & Treatment group & Control group & Treatment group &  \\
    \midrule
    Number of Links & 0.0977 & 0.7883 & 1.7842 & 4.2644 & 0.0282 \\  
    $\Delta$population (log scale) & 1.2210 & 1.1975 & 1.2210 & 1.1975 & -1.0359 \\
    $\Delta$GDP per capita (log scale) & 1.1717 & 1.1226 & 1.1717 & 1.1226 & -0.0814 \\
    Distance & 0.2628 & 0.2757 & 0.2628 & 0.2757 & 0.0004 \\
    \midrule
    Observations & 2928  & 12671 & 2928  & 12671 &  \\
    \bottomrule
    \bottomrule
    \end{tabular}%
  \label{tab:DID_apendix1}%
\end{table}%

\begin{table}[htbp]
  \centering
  \caption{Summary statistics of between East and West (1995-2005)}
    \begin{tabular}{lccccc}
          &       &       &       &       &  \\
    \midrule
    \midrule
    \multicolumn{1}{c}{\multirow{2}[3]{*}{West-East links}} & \multicolumn{2}{c}{Before} & \multicolumn{2}{c}{After} & \multirow{2}[3]{*}{Differences in Differences} \\
\cmidrule{2-5}          & Control group & Treatment group & Control group & Treatment group &  \\
    \midrule
    Number of Links & 0.1211 & 0.7786 & 1.8752 & 4.6832 & 0.0282 \\
    $\Delta$population (log scale) & 1.4371 & 1.3936 & 6.8174 & 6.7659 & -1.0359 \\
    $\Delta$GDP per capita (log scale) & 1.3754 & 1.3610 & 1.3124 & 1.3077 & -0.0814 \\
    Distance & 0.3760 & 0.3811 & 0.3760 & 0.3811 & 0.0004 \\
    \midrule
    Observations & 785   & 3270  & 785   & 3270  &  \\
    \bottomrule
    \bottomrule
    \end{tabular}%
  \label{tab:DID_apendix2}%
\end{table}%

\end{landscape}
\end{document}